\documentstyle[12pt]{article}

\def\R{ {\rm R \kern -.31cm I \kern .15cm}}
\def\C{ {\rm C \kern -.15cm \vrule width.5pt \kern .12cm}}
\def\Z{ {\rm Z \kern -.27cm \angle \kern .02cm}}
\def\N{ {\rm N \kern -.26cm \vrule width.4pt \kern .10cm}}
\def\1{{\rm 1\mskip-4.5mu l} }
\def\lsim{\raise0.3ex\hbox{$<$\kern-0.75em\raise-1.1ex\hbox{$\sim$}}}
\def\gsim{\raise0.3ex\hbox{$>$\kern-0.75em\raise-1.1ex\hbox{$\sim$}}}
\def\noi{\noindent}

\def\beq{\begin{equation}}   \def\eeq{\end{equation}}
\def\bea{\begin{eqnarray}}  \def\eea{\end{eqnarray}}
\def\nn{\nonumber}
\def\noi{\noindent}
\def\beeq{\begin{eqnarray}} \def\eeeq{\end{eqnarray}}
\newcommand\mysection{\setcounter{equation}{0}\section}
\renewcommand{\theequation}{\thesection.\arabic{equation}}
\newcounter{hran} \renewcommand{\thehran}{\thesection.\arabic{hran}}

\def\bmini{\setcounter{hran}{\value{equation}}
   \refstepcounter{hran}\setcounter{equation}{0}
   \renewcommand{\theequation}{\thehran\alph{equation}}\begin{eqnarray}}

\def\bminiG#1{\setcounter{hran}{\value{equation}}
\refstepcounter{hran}\setcounter{equation}{-1}
\renewcommand{\theequation}{\thehran\alph{equation}}
\refstepcounter{equation}\label{#1}\begin{eqnarray}}

\begin{document}
\centerline{\Large\bf Solvable Confining Gauge Theories at Large N}
\vskip 3 truecm

\centerline{\bf Ulrich ELLWANGER$^{\bf a}$, Nicol\'as WSCHEBOR$^{\bf
a,b}$}\vskip 3 truemm
\centerline{$^{\bf a}$ Laboratoire de Physique Th\'eorique
\footnote{Unit\'e Mixte de Recherche - CNRS - UMR 8627}}  
\centerline{Universit\'e de Paris XI, B\^atiment 210, F-91405 ORSAY
Cedex, France}

\vskip 1 truecm

\centerline{$^{\bf b}$ Institutos de F\'{\i}sica} 
\centerline{Facultad de Ciencias (Calle Igu\'a 4225, esq. Mataojo)} 
\centerline{and Facultad de Ingenier\'{\i}a (C.C.30, CP 1100),
Montevideo, Uruguay}

\vskip 5 truecm

\begin{abstract}

\noi In this letter we consider models with $N$ $U(1)$ gauge
fields $A_{\mu}^n$ together with $N$  Kalb-Ramond fields $B_{\mu
\nu}^n$ in the large N limit. These models can be solved explicitely
and exhibit confinement for a large class of bare actions. The
confining phase is characterized by an approximate ``low energy''
vector gauge symmetry under which the Kalb-Ramond fields
$B_{\mu\nu}^n$ transform. A duality transformation shows that
confinement is associated with magnetic monopoles condensation.

\end{abstract}

\vskip 1.5 truecm
\noi LPT Orsay 01-49 \par
\noi June 2001 \par

\newpage
\pagestyle{plain}
\baselineskip 18pt

\mysection{Introduction}
\hspace*{\parindent}
In various approaches towards a description of the confining phase of
gauge theories -- on the lattice, through duality with the Higgs
phase, ANO-strings or the "confining string" -- the introduction of a 
Kalb-Ramond fields $B_{\mu\nu}$ \cite{1r} as effective variables has
proved to be very useful [2 -- 10]. In the case of pure Yang Mills
theories they can be introduced as auxiliary fields for the abelian
components of the field strength [8,9] in the maximal abelian gauge 
\cite{11r} and, after performing the path integral over the
"non-diagonal" gauge fields (associated with non-diagonal generators),
one is left with an effective action involving abelian gauge fields and
Kalb-Ramond fields only.\par

Common to all these approaches is the idea that Kalb-Ramond fields are
effective variables only at low energy, i.e. in the infra-red regime.
It is thus natural to provide models with Kalb-Ramond fields as
effective variables with an ultra-violet cutoff, but to allow for
irrelevant operators in the corresponding bare action. On the other
hand it is sufficient to restrict oneself to abelian gauge theories (or
the abelian subsector of non-abelian theories): Monopole condensation,
which is believed to be the mechanism behind confinement [12, 13], is a
purely abelian phenomenon. In the context of non-abelian gauge theories
the relevance of the abelian sub-sector is conventially referred to as
"abelian dominance". \par

The purpose of the present paper is the study of models with $N$
abelian gauge fields $A_{\mu}^n$ and $N$ Kalb-Ramond fields $B_{\mu
\nu}^n$ in the large N limit. As we will see, they can be solved using
standard functional methods employed for large N field theories. The
emergence of a confining phase can be seen explicitely, and the
appearance of a ``low energy'' vector gauge symmetry allows for a
duality transformation showing that magnetic monopoles have condensed.
In this letter we present the essential results of this approach,
leaving many details (as the precise relation to Yang-Mills theories)
to a subsequent publication \cite{14r}. \par 

\mysection{The A$_{\mu}$ - B$_{\mu \nu}$ -- model}
\hspace*{\parindent} 

The starting point is the partition function for a model with the above
field content. Adding sources for $A_{\mu}^n$ and $B_{\mu \nu}^n$, $n =
1 \dots $ N, and a
covariant gauge fixing term the partition function reads

\beq
\label{1.1e}
e^{-W(J)} = {1 \over {\cal N}} \int {\cal D} A \ {\cal D}B \ 
e^{-S_{bare}(A,B) + \int
d^4x \left \{ {1 \over 2 \alpha} (\partial_{\mu}A_{\mu}^n)^2 + 
J_{A,\mu}^{n} A_{\mu}^n +
J_{B,\mu\nu}^{n}B_{\mu\nu}^n\right \} }\ . \eeq

Due to the N $U(1)$ gauge symmetries $S_{bare}(A,B)$ can only depend on
$F_{\mu \nu}$, hence we can write $S_{bare}(F,B)$. Next, in order to
allow for a large N expansion, we assume that $S_{bare}(F,B)$ depends
on $O(N)$ invariants (singlets) only. The aim would be to allow for a
dependence of $S_{bare}(F,B)$ on $O(N)$ singlets as general as
possible. Here we confine ourselves to the following ansatz: First we
introduce three Lorentz scalar $O(N)$ singlet operators

\bea
\label{1.2e}
{\cal O}_1(x) = \sum_{n=1}^N F_{\mu\nu}^n(x) F_{\mu\nu}^n (x) 
\quad , \nn \\
{\cal O}_2(x) = \sum_{n=1}^N F_{\mu\nu}^n(x) B_{\mu\nu}^n (x) 
\quad , \nn \\
{\cal O}_3(x) = \sum_{n=1}^N B_{\mu\nu}^n(x) B_{\mu\nu}^n (x) \quad .
\eea
 
\noi Then we take $S_{bare}(F,B)$ of the form

\beq
\label{1.3e}
S_{bare}(F,B) = \int d^4x \left \{ {\cal L}_{bare}({\cal O}_i) + {h 
\over 2} \left ( \partial_{\mu}  \widetilde{B}_{\mu \nu}^n \right )^2 +
{\sigma \over 2}  \left (  \partial_{\mu} B_{\mu \nu}^n \right )^2
\right \}
\eeq

\noi where

\beq
\label{1.4e}
\widetilde{B}_{\mu \nu}^n = {1 \over 2} \ \varepsilon_{\mu 
\nu\rho\sigma} \ B_{\rho\sigma}^n \quad .
\eeq

We allow ${\cal L}_{bare}$ in (\ref{1.3e}) to contain arbitrary
derivatives acting on the operators ${\cal O}_i$. This is still not the
most general form of $S_{bare}(F,B)$; one can certainly construct
infinitely many more $O(N)$ singlet operators which contain open
Lorentz indices and/or "internal" derivatives as the second and third
terms in (\ref{1.3e}). It can be argued \cite{14r}, however, that these
do not modify the essential features of the results obtained below.\par

In order to solve the model in the large $N$ limit we have to make 
assumptions on the $N$ dependence of the
parameters in $S_{bare}$. These assumptions can be summarized  by 
rewriting (\ref{1.3e}) as

\beq
\label{1.5e}
S_{bare}(F, B) = \int d^4x \left \{ N {\cal L}_{bare} \left ( 
{{\cal O}_i \over N}\right ) + {h \over 2}
\left ( \partial_{\mu} \widetilde{B}_{\mu\nu}^n \right )^2 + 
{\sigma \over 2} \left ( \partial_{\mu} 
B_{\mu \nu}^n \right )^2 \right \}
  \eeq

\noi where now the coefficients of ${\cal L}_{bare}$ are independent of
$N$. \par

\mysection{The large N solution}
\hspace*{\parindent} 

The most convenient formalism for the treatment of field  theories in
the large $N$ limit is the introduction of auxiliary fields for
composite $O(N)$ singlet operators \cite{15r}. In the present case we
introduce one auxiliary field $\phi_i$ for each of the bilinear $O(N)$ 
singlet operators ${\cal O}_i$ in eq. (\ref{1.2e}). This amounts to
re-write the term involving $N {\cal L}_{bare} \left ({\cal O}_i /
N\right )$ in the exponent of (\ref{1.1e}) as 

\beq
\label{2.1e}
e^{-N\int d^4x{\cal L}_{bare}\left ( {{\cal O}_i \over N} \right )} = 
{1 \over {\cal N}} \int {\cal D} \phi_i \ e^{-NG_{bare}(\phi_i)
- \int d^4x\phi_i {\cal O}_i} \quad . \eeq

\noi In the large $N$ limit the path integral on the right-hand side of
(\ref{2.1e}) can  be replaced by its stationary point, and the relation
between $G_{bare}$ and ${\cal L}_{bare}$ becomes

\beq
\label{2.2e}
N \int d^4x\ {\cal L}_{bare} \left ( {{\cal O}_i \over N}\right ) = N 
G_{bare}(\phi_i) + \int d^4x \ \phi_i \
{\cal O}_i \quad . \eeq

\noi Equation (\ref{2.2e}) allows, in principle, to construct 
$G_{bare}(\phi_i)$ from ${\cal L}_{bare}$, although here we allow 
${\cal L}_{bare}$ to be an arbitrary functional (including derivatives)
of ${\cal O}_i$. Next we insert eq. (\ref{2.1e}) into (\ref{1.1e}),
which becomes

\bea
\label{2.3e}
&&e^{-W(J)} = {1 \over {\cal N}} \int {\cal D} \phi_i \int {\cal 
D}A \ {\cal D}B \nn \\
&& \times e^{-NG_{bare}(\phi_i)  - \int d^4x \left \{ 
\phi_i {\cal O}_i + {h \over 2} (\partial_{\mu}\tilde{B}_{\mu\nu}^n)^2
+ {\sigma \over 2}  (\partial_{\mu}B_{\mu\nu}^n)^2 + {1 \over 2 \alpha}
(\partial_{\mu} A_{\mu}^n)^2 - J_{A,\mu}^{n}A_{\mu}^n - 
J_{B,\mu\nu}^{n} B_{\mu\nu}^n \right \} } \nn \\
\eea

\noi The ${\cal D}A \ {\cal D}B$ path integrals have become Gaussian
in (\ref{2.3e}). In order to express the result in compact form we
introduce the notation $\varphi_r^n = \left \{ A_{\mu}^n , B_{\mu
\nu}^n \right \}$, i.e. the indices $r$ attached to the fields
$\varphi^n$  denote both the different fields $A^n$, $B^n$ and the
different Lorentz indices. Correspondingly we introduce notation
$J_r^n$ for $\{J_{A,\mu}^n, J_{B,\mu\nu}^n\}$. The
result of the Gaussian integration over ${\cal D}A \ {\cal D}B$ can
now be written as

\beq
\label{2.4e}
e^{-W(J)} = {1 \over {\cal N}} \int {\cal D} \phi_i \ 
e^{-NG_{bare}(\phi_i) - N\Delta G(\phi_i) + {1 \over 2}
\int d^4x_1 d^4x_2 \left \{ J_r^n(x_1) P^{rs}(x_1, x_2, \phi_i) 
J_s^n(x_2) \right \} } \eeq

\noi with

\beq
\label{2.5e}
\Delta G(\phi_i) = - {1 \over 2} \ Tr \log \left ( P^{rs}(x_1, x_2, 
\phi_i ) \right ) \quad .
\eeq

The propagators $P^{rs}$ of the $A_{\mu}^n$, $B_{\mu\nu}^n$ -- system
are proportional to $\delta_{n,m}$ with $n, m = 1 \dots N$ and we took
care of the resulting contribution from the trace in (\ref{2.5e}) by
the  explicit factor $N$ multiplying $\Delta G$ in (\ref{2.4e}). The
propagators $P_{\mu,\nu}^{AA}$,  $P_{\mu,\rho\sigma}^{AB}$ and
$P_{\mu\nu,\rho \sigma}^{BB}$ depend on the terms  $\phi_i {\cal 
O}_i$, ${h \over 2} (\partial \widetilde{B})^2$, ${\sigma \over 2}
(\partial B)^2$ and ${1 \over 2\alpha} (\partial A)^2$ in the exponent
of (\ref{2.3e}). Simple explicit expressions can be obtained only for
constant fields $\phi_i$; in this case one finds for $\Delta G$
(in the Landau gauge $\alpha \to 0$)

\beq
\label{2.6e}
\Delta G(\phi_i) = {3 \over 2} \int d^4x \int {d^4p \over ( 2 \pi  )^4}
\Big [ \log ( \phi_1 \sigma p^2 + 4 \phi_1  \phi_3  - \phi_2^2  )  +
\log \left ( hp^2 + 4 \phi_3 \right  ) \Big ] \ .  \eeq

\noi The $d^4p$ integral in (\ref{2.6e}) has to be performed with 
an UV cutoff $\Lambda^2$. The result simplifies considerably if one
introduces

\beq
\label{2.7e}
\Sigma = {4 \phi_1 \phi_3 - \phi_2^2 \over \sigma \phi_1} \quad .
\eeq

\noi Up to field independent terms one then obtains

\bea
\label{2.8e}
\Delta G(\phi_i) = &&{3 \over 32\pi^2} \int d^4x \left [ \left ( 
\Lambda^4 - \Sigma^2 \right ) \log \left (\sigma 
\phi_1 (\Lambda^2  + \Sigma) \right )
+ \Sigma^2 
\log \left ( \sigma \phi_1 \Sigma\right ) 
+ \Lambda^2 \Sigma
\right . 
\nn \\
&&\left . + \left ( \Lambda^4 - {16 \phi_3^2 \over h^2} \right ) 
\log \left ( \Lambda^2 h + 4 \phi_3 \right ) 
+ {16 \phi_3^2 \over h^2} \log \left ( 4 \phi_3 \right )
  + 4 \Lambda^2{\phi_3 \over h} \right ] .  
\eea

\noi This expression for $\Delta G(\phi_i)$ has to be inserted into 
(\ref{2.4e}) and, in the large N limit, the ${\cal D}\phi_i$ path
integral is again dominated by its stationary point(s).  Hence $W(J)$
becomes

\beq
\label{2.9e}
W(J) = N G(\widehat{\phi}_i) - {1 \over 2} \int d^4x_1 d^4x_2 
\left \{ J_r^n(x_1) \ P^{rs}(x_1, x_2,
\widehat{\phi}_i)J_s^n (x_2) \right \} \eeq

\noi where

\beq
\label{2.10e}
G(\phi_i) = G_{bare} (\phi_i) + \Delta G(\phi_i)
\eeq

\noi and $\widehat{\phi}_i \equiv \widehat{\phi}_i(J)$ satisfy the 
three equations (recall $i = 1,2,3$)

\beq
\label{2.11e}
\left [ {\delta \over \delta \phi_i} \Big (NG(\phi_i) - {1 \over 2}
\int  d^4x_1 d^4x_2 \ J_r^n(x_1) \ P^{rs}(x_1, x_2, \phi_i) J_s^n
(x_2) \Big ) \right ]_{\widehat{\phi}_i(J)} = 0 \ . \eeq

\noi The model is thus solved, for given $G_{bare} (\phi_i)$, up to
the technical problem of finding the stationary points
$\widehat{\phi}_i(J)$. \par

Next we wish to show that the particular configuration where

\beq
\label{2.12e}
4 \widehat{\phi}_1 \widehat{\phi}_3 - \widehat{\phi}_2^2 = 0 \quad .
\eeq

\noi (or $\widehat{\Sigma} = 0$) is a "natural" solution of the three
stationary point equations (\ref{2.11e}), i.e. a solution which
requires no fine tuning of the parameters in $G_{bare} (\phi_i)$. Below
we will see that the phase where (\ref{2.12e}) holds is the confining
phase of the model. \par

In order to "see" the solution (\ref{2.12e}) of the eqs. (\ref{2.11e})
it is necessary to regularize the singularity of the derivatives of
$\Delta G(\phi_i)$ w.r.t. the fields at $\Sigma = 0$. The origin of the
non-analytic behaviour of $\Delta G(\phi_i)$ at $\Sigma = 0$ is the
infrared behaviour of the propagators $P^{r s}(\phi_i)$ which, in
momentum space, behave like $P^{r s}(q^2, \phi_i) \sim q^{-4}$ for
$\Sigma = 0$. In order to regularize these infrared singularities we
perform the $d^4 p$ integral in (\ref{2.6e}) also with an infrared
cutoff $k^2$. For $J^n_r = 0$ (hence we write $\widehat{\phi}_i^0$
instead of $\widehat{\phi}_i$) the three stationary point equations
(\ref{2.11e}) can then be brought into the form

\bea
\label{2.13e}
\left [ {\delta G_{bare} \over \delta \phi_1} + {3 \over 32 \pi^2 
\phi_1} \left ( \Lambda^4 - k^4 \right )
\right ]_{\widehat{\phi}_i^0} = 0 \ , 
\nn \\ 
\left [ {\delta G_{bare} \over \delta \phi_3} + {3 \over 4 \pi^2 h^2} 
\left \{ 4 \phi_3 \log \left (
{4 \phi_3 + hk^2 \over 4 \phi_3 + h \Lambda^2} \right ) + h \left ( 
\Lambda^2 - k^2 \right ) \right \} \right
]_{\widehat{\phi}_i^0} = 0 \ , 
\nn \\ 
\left [ {\delta G_{bare} \over \delta \Sigma} + {3 \over 16 \pi^2 } 
\left \{ \Sigma \log \left (
{\Sigma + k^2 \over \Sigma + \Lambda^2} \right ) +  
\Lambda^2 - k^2 \right \} \right
]_{\widehat{\phi}_i^0} = 0 \ .
\eea

\noi The solutions ${\widehat{\phi}_i^0}$ of (\ref{2.13e}), and hence 
${\widehat{\Sigma}^0}$, depend on the infrared cutoff $k^2$. Let us now
assume that

\beq
\label{2.14e}
\left. - \ {\delta G_{bare} \over \delta \Sigma} 
\right |_{\widehat{\Sigma}^0 = 0}  -{3 \over 16 \pi^2} \ \Lambda^2 > 0
 \quad . 
\eeq

\noi Then the last of the  stationary point equations (\ref{2.13e})
implies ${\widehat{\Sigma}^0}(k^2) < 0$ for $\Lambda^2$~$>$~$k^2 > 0$, and
${\widehat{\Sigma}^0}(k^2)$ behaves as follows for $k^2 \to 0$:

\bea
\label{2.15e}
\widehat{\Sigma}^0(k^2) \to 0_{-\varepsilon} \quad , \quad
\left. {3 \over 16 \pi^2} \widehat{\Sigma}^0 \log \left (
{\widehat{\Sigma}^0 + k^2 \over  \widehat{\Sigma}^0 + \Lambda^2}\right )
\ \to  \ - \ {\delta G_{bare} \over \delta \Sigma} 
\right |_{\widehat{\Sigma}^0 = 0}  -{3 \over 16 \pi^2} \ \Lambda^2 
\ . \eea

\noi Hence, under the condition (\ref{2.14e}), we obtain
$\widehat{\Sigma}(0) = 0$ naturally. Note that this stationary point
would not have been observed if one puts $k^2 = 0$ from the start. \par

What is the meaning of the condition (\ref{2.14e}) or, better, 
${\widehat{\Sigma}^0}(k^2) < 0$? To this end we push the infrared
cutoff $k^2$ upwards until it reaches the UV cutoff $\Lambda^2$, and
investigate the consequence of ${\widehat{\Sigma}^0}(\Lambda^2) < 0$ .
Given the definition (\ref{2.7e}) for $\Sigma$, and for
$\widehat{\phi}_1^0(\Lambda^2) > 0$, this latter condition reads 

\beq
\label{2.16e}
4 \widehat{\phi}_1^0 (\Lambda^2) \widehat{\phi}_3^0 
(\Lambda^2) - (\widehat{\phi}_2^0 (\Lambda^2))^2< 0  \quad .
\eeq

\noi For $k^2 = \Lambda^2$ the contribution $\Delta G(\phi_i)$ to 
$G(\phi_i)$ in (\ref{2.10e}) vanishes, and the configurations 
$\widehat{\phi}_i^0(\Lambda^2)$ are the stationary points of $G_{bare}
(\phi_i)$. From the Legendre transformation (\ref{2.2e}) and the
definitions  (\ref{1.2e}) of the operators ${\cal O}_i$ it is now
straightforward to see that the inequality (\ref{2.16e})
implies (with  $\varphi_r^n$ as below (\ref{2.3e}))

\beq
\label{2.17e}
\left. Det\left({\delta^2 {\cal L}_{bare} \over  \delta \varphi_r^n
\delta \varphi_s^n } \right )\right |_{F=B=0} < 0 \ .
\eeq

\noi (\ref{2.17e}) corresponds to a ${\cal
L}_{bare}(F,B)$ which is {\it non-convex} at the origin of field space.
\par

All this is very similar to the case of a non-convex bare scalar
potential: The effective scalar potential has to be semi-convex, and if
the bare scalar potential is sufficiently non-convex the effective
potential  becomes flat in its "inner" region. As in the present case
the observation of this phenomenon requires the introduction of an
"artificial" infrared cutoff $k^2$, and a careful discussion of the
limit $k^2 \to 0$ \cite{16r}. We emphasize that consequently the
emergence of the "confining" phase (\ref{2.12e}) (see below) is not a
particular feature of the large N limit; the advantage of the large N 
limit is only to allow for an explicit study of this phenomen (for
given parametrizations of ${\cal L}_{bare}(F,B)$ or
$G_{bare}(\phi_i)$).

\mysection{Properties of the confining phase}
\hspace*{\parindent} 
In the following we assume that the necessary inequality on the
parameters of ${\cal L}_{bare}(F,B)$ (a sufficiently negative curvature
at the origin of field space) for the reach of the confining phase
(\ref{2.12e}) is satisfied. In order to discuss its properties it is
more convenient to switch from $W(J)$ in (\ref{2.9e}) to the effective
action $\Gamma (A, B)$ via a Legendre transform with respect to the
sources $J$:

\beq
\label{3.1e}
\Gamma (A,B) = W(J) + \int d^4x \left ( J_{A, \mu}^n A_{\mu}^n 
+ J_{B, \mu \nu}^n B_{\mu \nu}^n \right
) \quad . \eeq

\noi As discussed in detail in \cite{14r} one obtains

\beq
\label{3.2e}
\Gamma(A,B) =N G(\widehat{\phi}_i) + \int d^4x \left 
( \widehat{\phi}_i {\cal O}_i + {h \over
2} \left ( \partial_{\mu} \widetilde{B}_{\mu\nu}^n \right )^2 + 
{\sigma \over 2} \left ( \partial_{\mu}
B_{\mu\nu}^n \right )^2 \right ) \eeq

\noi with $G(\widehat{\phi}_i)$ as in (\ref{2.10e}), and the stationary
point equations (\ref{2.11e}) for $\widehat{\phi}_i$ can be written as

\beq
\label{3.3e}
\left [ \ {\delta \Gamma\over \delta \phi_i} \ 
\right ]_{\widehat{\phi}_i(A,B)} = 0 \quad .
\eeq

First we note that because of the relation (\ref{2.12e}) the expression 
$\widehat{\phi}_i {\cal O}_i$ in (\ref{3.2e}) becomes

\beq
\label{3.4e}
\widehat{\phi}_i {\cal O}_i = \sum_n \left ( 
\sqrt{\widehat{\phi}_1} \ F_{\mu \nu}^n +
\sqrt{\widehat{\phi}_3} \ B_{\mu \nu}^n \right )^2 \quad . \eeq

\noi Consequently it is invariant under the following gauge symmetry
involving vector-like gauge parameters $\Lambda_{\mu}^n$ \cite{1r}:

\bea
\label{3.5e}
&&\delta A_{\mu}^n(x) = \Lambda_{\mu}^n(x) \ , \ \delta 
F_{\mu\nu}^n(x) = \partial_{\mu} \Lambda_{\nu}^n (x) - \partial_{\nu}
 \Lambda_{\mu}^n(x) \equiv \Lambda_{\mu\nu}^n  (x) \ , \nn \\
&&\delta B_{\mu}^n(x) = \sqrt{{\widehat{\phi}_1 \over 
\widehat{\phi}_3}} \ \Lambda_{\mu\nu}^n (x) \quad .\eea

\noi In addition one finds that the term  $\sim (\partial_{\mu}
\widetilde{B}_{\mu\nu}^n)^2$ in (\ref{3.2e}) is also invariant under
(\ref{3.5e}) thanks to a Bianchi identity, provided the configurations 
$\widehat{\phi}_i$ are constant in $x$. (Note that, from eq. 
(\ref{3.3e}) with $\Gamma$ as in (\ref{3.2e}), constant configurations
$\widehat{\phi}_i$ result from constant configurations ${\cal O}_i$;
however, from their definition (\ref{1.2e}), constant ${\cal O}_i$ do
{\it not} necessarily imply constant configurations $F_{\mu \nu}^n$ and
$B_{\mu \nu}^n$.) The last term $\sim (\partial_{\mu} B_{\mu\nu}^n)^2$
in (\ref{3.2e}) behaves as a gauge fixing term of the symmetry
(\ref{3.5e}), and its presence insures the existence of the
propagators. \par

It is to be expected that the symmetry (\ref{3.5e}) is broken by 
higher derivative terms (beyond the gauge fixing term): The bare action
$S_{bare}$ (\ref{1.5e}) of the model  does certainly not exhibit the
symmetry (\ref{3.5e}), and the  Green functions at large
non-exceptional Euclidean momenta with $p^2 \to \Lambda^2$ are
generated by  $S_{bare}$. This fact is realized by the dependence of
the effective action on higher derivative terms. The symmetry
(\ref{3.5e}) is thus a pure ``low energy'' phenomenon. The implication
of the gauge symmetry (\ref{3.5e}) on modes of the $U(1)$ gauge fields
$A_{\mu}^n$ which correspond to constant configurations  ${\cal O}_i$
is that they can be ``gauged away'' and ``eaten'' by the (massive or
even infinitely massive) Kalb-Ramond fields $B_{\mu\nu}^n$, in some
analogy to the ordinary Higgs effect \cite{1r}.\par

In addition the symmetry (\ref{3.5e}) allows for a duality
transformation: A priori the dual of a $U(1)$ gauge field $A_{\mu}^n$
(in $d=4$) is again a $U(1)$ gauge field $C_{\mu}^n$ (whose field
strength  tensor will be denoted by $F_{\mu\nu}^{c,n}$), and the dual
of a Kalb-Ramond field $B_{\mu\nu}^n$ is a (pseudo-) scalar
$\varphi^n$. In the present case the duality transformations mix the
fields and read

\bea
\label{3.6e}
{1 \over 2}\ F_{\mu\nu}^{c,n} &=& \widehat{\phi}_1  
\widetilde{F}_{\mu\nu}^n + \sqrt{\widehat{\phi}_1 
\widehat{\phi}_3} \widetilde{B}_{\mu\nu}^n \quad , \nn \\
\partial_{\mu} \varphi^n + C_{\mu}^n &=& {h \over 2} \ 
\sqrt{{\widehat{\phi}_1 \over \widehat{\phi}_3}} \  
\partial_{\nu} \widetilde{B}_{\nu\mu}^n
\eea

\noi where the tildes on $\widetilde{B}_{\mu\nu}^n$ and 
$\widetilde{F}_{\mu\nu}^n$ have been defined in (\ref{1.4e}). The
corresponding dual action reads (as obtained from (\ref{3.2e}) without
$N G(\widehat{\phi}_i)$ and without the "gauge fixing" term)

\beq
\label{3.7e}
\Gamma_{Dual}(C, \varphi) = \int d^4x \left \{ {1 \over 4
\widehat{\phi}_1 } F_{\mu\nu}^{c,n}  F_{\mu\nu}^{c,n} + {2 \over h}
{\widehat{\phi}_3 \over  \widehat{\phi}_1} \left ( \partial_{\mu}
\varphi^n + C_{\mu}^n \right )^2 \right \} \ . \eeq

Note that, due to the implicit dependence of $\widehat{\phi}_i$ on $F$
and $B$, these duality transformation are non-linear. Actually one
finds \cite{14r} that only half of the equations of motion and Bianchi
identities are exactly interchanged through (\ref{3.6e}) and
(\ref{3.7e}), whereas the other half holds again only for constant
configurations $\widehat{\phi}_i$ and hence ${\cal O}_i$. Thus duality
is realized at the non-linear level again only in the corresponding
"low energy" regime. \par

The physical interpretation of the dual action (\ref{3.7e}) is 
obviously the one of an abelian $U(1)^N$ Higgs model in the
spontaneously broken phase where $\varphi^n$  represent the Goldstone
bosons, and where the gauge fields $C_{\mu}^n$ have acquired a mass 
$2(\widehat{\phi}_3/h)^{1/2}$. Since this represents the ``low energy
effective action'' of a theory in which the ``dual''  electric charge
has condensed in the vacuum, the original action (\ref{3.2e}) with
(\ref{3.4e}) corresponds to the situation where the ``magnetic''
charge has condensed in the vacuum. \par

Let us turn to the response of the model in the confining phase with
respect to external sources. The expression for $W(J)$ has been given
in eq. (\ref{2.9e}) in the preceeding section, and first we concentrate
ourselves on the term quadratic in the sources $J_r^n$. Let us start
with a source $J_{A,\mu}^n(x)$ for the fields  $A_{\mu}^n$ only, which
is of the form of a Wilson loop: 

\beq
\label{3.8e}
J_{A,\mu}^n(x) = ig_A \int_C dx'_{\mu} \ \delta^4(x - x') \quad .
\eeq

\noi The term quadratic in $J$ in (\ref{2.9e}) then becomes

\beq
\label{3.9e}
{Ng_A^2 \over 2} \int_C dx_{1,\mu} \int_C dx_{2,\nu} \ 
P_{\mu,\nu}^{AA}(x_1 - x_2)
\eeq

\noi where $P_{\mu,\nu}^{AA}$ has to be obtained from the action 
(\ref{3.2e}) with (\ref{3.4e}):

\bea
\label{3.10e}
P_{\mu, \nu}^{AA}(z) =&& {1 \over 16 \pi^2\widehat{\phi}_1} \left \{ 
\delta_{\mu\nu} \left ( {1 \over |z|^2} - {2\widehat{\phi}_3 \over
\sigma} \log |z| + \hbox{const.}  \right ) \right .\nn \\
&&\left . - {1 \over 2} \partial_{\mu}\partial_{\nu} \left ( \log 
|z| - {\widehat{\phi}_3 |z|^2\over 2\sigma} (\log
|z| + \hbox{const.'}) \right ) \right \} \ .
\eea 

\noi The (actually divergent) constants in (\ref{3.10e}) disappear in
the expression (\ref{3.9e}). In the limit where the (minimal) surface
$S$ enclosed by the loop $C$ in (\ref{3.8e}) becomes very large one
finds that the expression (\ref{3.9e}) is proportional to $S$, thus 
one obtains the area law for the expectation value of the Wilson
loop. \par

However, at first sight an inconsistency arises due to the long-range
behaviour of the propagator $P_{\mu, \nu}^{AA}(z)$ in (\ref{3.10e}):
Let us imagine that space-time is filled with "virtual" Wilson loops
(originating, e.g., from vacuum bubbles of virtual quark-antiquark
pairs), and let us compute the corresponding contribution to the action
due to the interactions among different "virtual" Wilson loops. Even if
one assumes that these "virtual" Wilson loops are arbitraryly tiny in
size, localized in space-time and if one averages over their
orientation in space-time, the contribution to the action induced by
the long-range behaviour of the propagator $P_{\mu, \nu}^{AA}(z)$ in
(\ref{3.10e}) diverges (logarithmically) in the infinite volume limit.
\par

This infinity can be avoided, however, once one realizes that {\it all}
components $\{r,s\} = \{A,B\}$ of the propagators $P^{rs}(z)$ in
(\ref{2.9e}) have a "bad" long range behaviour. The precise expressions
for all propagators, as obtained from the action (\ref{3.2e}), will be
given in \cite{14r}. One finds that all terms in the propagators which
decrease not sufficiently fast at infinity in order to avoid the above
infrared divergence (which originate from $q^{-4}$-terms in momentum
space) cancel in the sum over $r$ and $s$ in $J_r^n P^{rs} J_s^n$ in
(\ref{2.9e}) if the sources $J_{A,\mu}^n$ and $J_{B,\mu\nu}^n$ satisfy

\beq
\label{3.11e}
\sqrt{\widehat{\phi}_3} \ J_{A,\mu}^n(x) = 2 \sqrt{\widehat{\phi}_1}\
\partial_{\nu}   J_{B,\nu\mu}^n(x) \ .
\eeq

This result can also be phrased as follows: In the presence of an
arbitrary background of "virtual" Wilson loops it costs an infinite
amount of action (or energy) to "switch on" sources $J_{A,\mu}^n$
and/or $J_{B,\mu\nu}^n$ which are {\it not} related as in
(\ref{3.11e}), due to the interactions induced between the sources and
the background of "virtual" Wilson loops induced by the long-range
terms in the propagators. \par

If the source $J_{A,\mu}^n$ is of the form of a Wilson loop (\ref{3.8e})
one finds that (\ref{3.11e}) implies that the source $J_{B,\nu\mu}^n$ is
of the form of a "Wilson surface",

\beq
\label{3.12e}
J_{B,\mu\nu}^n(x) = ig_B \int_S d^2\sigma_{\mu\nu}(z) \ \delta^4(x - z)
\quad ,
\eeq

\noi where the surface $S$ is bounded by the loop $C$ in (\ref{3.8e})
(but otherwise arbitrary) and where $g_B$ satisfies 
$\sqrt{\widehat{\phi}_1} \ g_B = \sqrt{\widehat{\phi}_3} \ g_A/2$.\par

It is straightforward to see that the condition (\ref{3.11e}) on the
sources is equivalent to the condition that the couplings $J_{A,\mu}^n
A_{\mu}^n + J_{B,\mu\nu}^n B_{\mu\nu}^n$ of the fields to the sources
respect the "low energy" gauge symmetries (\ref{3.5e}). In the case of
conventional gauge symmetries these conditions can (and have to) be
imposed by hand in order to ensure renormalizability and unitarity of
the theory. In the present model, on the one hand, they cannot be
imposed from the beginning, since the associated (low energy) gauge
symmetries appear only at the level of the effective action once the 
equations of motion of the fields $\phi_i$ are satisfied. Although
renormalizability is not an issue here, since we consider an effective
low energy theory with a fixed UV cutoff, it is interesting to see that
the corresponding condition on the sources is generated dynamically in
the sense that its violation costs infinite action. \par

Clearly we now have to reconsider the expectation value of the Wilson
loop, which consists now of a source (\ref{3.8e}) for $A_{\mu}^n$ {\it
and} a source (\ref{3.12e}) for $B_{\mu\nu}^n$ inserted into the term
quadratic in $J$ in (\ref{2.9e}) (higher orders in $J$ will be discussed
below). Using all propagators $P^{rs}$ from \cite{14r} this term
becomes

\bea
\label{3.13e}
&&\int_S d^2\sigma_{\mu\nu}(z_1) \int_S d^2\sigma_{\rho\sigma}(z_2) 
{-g_F^2 \over 16 \pi^2 \widehat{\phi}_1} \sqrt{{\widehat{\phi}_3 \over
h}}  \nn \\
&& \times \left ( T_{1,\mu\nu,\rho\sigma}(\partial ) - {4 \phi_3 \over
h} T_{2,\mu\nu ,\rho\sigma} \right )  {1 \over |z_1 - z_2|} K_1\left (
2 |z_1 - z_2| \sqrt{{\widehat{\phi}_3 \over  h}} \right )   
\eea

\noi with

\bea
\label{3.14e}
&&T_{1, \mu\nu , \rho \sigma}(\partial) = \delta_{\mu \rho}
\partial_{\nu}  \partial_{\sigma} - \delta_{\mu \sigma} \partial_{\nu}
\partial_{\rho} - \delta_{\nu \rho} \ \partial_{\mu} \partial_{\sigma}
+ \delta_{\nu  \sigma} \ \partial_{\mu} \partial_{\rho} \ , \nn \\
&&T_{2, \mu\nu , \rho \sigma} = \delta_{\mu \rho} \ \delta_{\nu 
\sigma} - \delta_{\mu \sigma} \ \delta_{\nu
\rho} \ ,
\eea

\noi and where $K_1$ is a Bessel function. In the limit where the
surface $S$ becomes large the expression (\ref{3.13e}) behaves as
 
\beq
\label{3.15e}
S \cdot {2g_F^2 \over \pi \widehat{\phi}_1} \left ( {\widehat{\phi}_3
\over h}\right  )^{3/2} \int_0^{\infty} dz \ K_1\left (
2z\sqrt{{\widehat{\phi}_3 \over h}} \right )\quad . \eeq

\noi Hence it implies the area law in spite of the cancellations of
the  long range contributions of the propagators. (Since we had omitted
the UV cutoff in the space-time propagators the expression
(\ref{3.15e}) is seemingly UV divergent). \par

The preceeding results, based on a treatment of the term quadratic in
$J$ in (\ref{2.9e}), have obvious interpretations in the context of the
stochastic vacuum model \cite{17r} for Yang-Mills theories: There, in
the Gaussian approximation, the expectation value of the Wilson loop is
given by the expectation value of the field strength correlator, which
plays the same role as the term quadratic in $J$ in (\ref{2.9e}). At
first sight an ambiguity appears: A priori it is not clear, whether the
Yang-Mills Wilson loop reappears in our "effective low energy model"
(after integrating out the off-diagonal gluons in the MAG, see
\cite{14r}) in the form of a source $J_A$, $J_B$ or, most likely, as a
combination of both (or even in the form of additional terms in
$S_{bare}$). The condition (\ref{3.11e}) fixes this ambiguity.\par

If we identify naively the term quadratic in $J$ in (\ref{2.9e}) with
the expectation value of the field strength correlator, the area law
obtained from (\ref{3.9e}) above corresponds to a function $D_1(x^2)$
in the standard decomposition of the field strength correlator
\cite{17r} which decreases only as $|x|^{-2}$ for large $|x|$. Such a
behaviour is strongly disfavoured by lattice measurements \cite{18r}
of the Yang-Mills field strength correlator. On the other hand the
results (\ref{3.13e}) and (\ref{3.15e}), which are the consequence of
the condition (\ref{3.11e}), agree well with with the lattice
measurements of the Yang-Mills field strength correlator
\cite{18r} and are in fact identical to the results for this correlator
obtained in various models \cite{10r,19r,20r}. \par

We recall, however, that up to now we have only discussed the term
quadratic in $J$ in (\ref{2.9e}). Let us note that this term coincides
(up to an irrelevant constant) with $W(J)$ to ${\cal O}(J^2)$, provided
that we replace $\widehat{\phi}_i$ by $\widehat{\phi}_i^0$ in
$P^{rs}(\widehat{\phi}_i)$: From the stationary point equations
(\ref{2.11e}) for $\widehat{\phi}_i$ we have  $\widehat{\phi}_i(J) =
\widehat{\phi}_i^0 + {\cal O}(J^2)$, and -- since the
$\widehat{\phi}_i^0$ are stationary points of $NG(\phi_i)$ -- we thus
have $NG(\widehat{\phi}_i(J)) = NG(\widehat{\phi}_i^0) + {\cal
O}(J^4)$. Beyond an expansion in powers of $J$ the stationary point
equations (\ref{2.11e}) are, however, cumbersome to solve since they
involve the full propagators $P^{rs}(\phi_i)$. \par

To this end it is wiser to start with the effective action $\Gamma (A,
B)$ as given in eq. (\ref{3.2e}). One should to solve the combined
equations of motion for $A_{\mu}^n$, $B_{\mu\nu}^n$ in the presence of
sources $J_{A,\mu}^n$ and $J_{B,\mu\nu}^n$ together with the equations
(\ref{3.3e}) for $\phi_i$ (in some analogy with the approach in
\cite{19r} based on the dual abelian Higgs model). The corresponding
solutions have to be inserted into $\Gamma (A,B)$ in (\ref{3.2e}), and
then one has to ``undo'' the Legendre transformation (\ref{3.1e}) in
order to obtain $W(J)$. This last step is actually trivial since
$\Gamma (A,B)$ is quadratic in $A$, $B$: It suffices to change the sign
of the second term in the expression (\ref{3.2e}). Then one can study
the dependence of $W(J)$ on $J$ in its full beauty. \par

As discussed in somewhat more detail in \cite{14r} the effect of such a
more complete calculation can be estimated in the simple case where the
sources $J$ are non-vanishing only inside a finite volume $V$ (to be
identified with a "Wilson surface" of finite width $\Delta L$), and
where derivative terms in $G(\phi_i)$ are neglected: Then the solutions
$\widehat{\phi}_i(x)$ of the stationary point equations differ from 
$\widehat{\phi}_i^0$ inside $V$, but coincide with $\widehat{\phi}_i^0$
outside $V$. Hence one obtains an additional contribution to the action
proportional to $V$, or a contribution to the energy of the
configuration proportional to the diameter of the Wilson surface (at
fixed time $t$). This picture supports "flux-tube" models for the
origin of the string tension. Its details (such as the shape of the
fields $\widehat{\phi}_i$ perpendicular to the string axis) depend,
however, on the precise form of $G_{bare}(\phi_i)$ and hence of ${\cal
L}_{bare}({\cal O}_i)$. (At this point it is useful to note that the
auxiliary fields $\phi_i$, as introduced in (\ref{2.1e}), parametrize
vevs of the bilinear operators ${\cal O}_i$ in (\ref{1.2e}).) \par

We close this section with a comment on the physical spectrum of the
model. In momentum space the propagators $P^{rs}(q^2)$ for the
$A_{\mu}$-$B_{\mu \nu}$ -- system have, in the confining phase,
$q^{-4}$ singularities for $q^2 \to 0$ as well as poles at $q^2 = 
-4\widehat{\phi}_3/h$. The $q^{-4}$ singularities for $q^2 \to 0$ do
not correspond to asymptotic physical states. The poles at $q^2 = 
-4\widehat{\phi}_3/h$, on the other hand, would disappear if we would
replace the constant $h$ in the ansatz (\ref{1.3e}) for $S_{bare}$ by a
function $h(q^2)$ such that $h(q^2)$ vanishes sufficiently rapidly for
large $|q^2|$. This would not modify any of our essential results, but
would be motivated by the idea that the Kalb-Ramond fields $B_{\mu
\nu}^n$ have originally be introduced into a more "microscopic" theory
(as a Yang-Mills theory) as auxiliary fields (for, e.g., the abelian
field strengths $F_{\mu \nu}^n$ \cite{8r,9r,14r}), and their kinetic
terms are thus loop-effects of modes which have been integrated out (as
the off-diagonal gluons). Then none of the degrees of freedom of the 
$A_{\mu}$-$B_{\mu \nu}$ -- system would appear as asymptotic states,
consistent with the absence of ($N_c-1$) -- plets in a $SU(N_c)$
Yang-Mills theory. \par

We have also searched for "bound states", i.e. poles in the 
propagators of the $\phi_i$ fields. Such poles are not present,
essentially because bosonic loop contributions to kinetic terms of
auxiliary $O(N)$ singlet fields differ in sign with respect to
fermionic loop contributions (which {\it do} generate propagating bound
states \cite{15r}). One may be deceived by the absence of 
``glueballs'', if the present model is interpreted as an effective low
energy theory for $SU(N_c)$ Yang-Mills theory. However, we recall that
the present model would only describe the abelian subsector of
$SU(N_c)$ Yang-Mills theory in the MAG \cite{8r,9r,14r}, and that we
finally have to add the off-diagonal gluons as well. Our theory induces
confining interactions among all fields which couple $A_\mu^n$ and
$B_{\mu \nu}^n$, hence the off-diagonal gluons will necessarily form
bound states which will correspond to the desired glueballs. \par

Thus, if we use functions $h(q^2)$ with the above-mentioned properties
in the ansatz (\ref{1.3e}) for $S_{bare}$, the model has no asymptotic
states at all. Its only "meaning" is then to react to external sources,
and to confine them as discussed before. \par

\mysection{Discussion and Outlook}

We have studied a class of four-dimensional $U(1)$ gauge theories
including Kalb-Ramond fields, which exhibit confinement and allow
nevertheless -- in the large N limit  -- for controllable computations
in the infrared regime. Some features of the confining phase correspond
quite to our expectations, notably the possiblity to perform a duality
transformation of the low energy part of the effective action and thus
to interpret confinement as monopole condensation. A technically
related phenomenon is the appearance of a low energy vector gauge
symmetry, which allows to ``gauge away'' the low momentum modes of the
abelian gauge fields  $A_{\mu}^n$. \par

An interesting feature is the origin of the relation among the
parameters of the effective action which generates the above symmetry:
The bare action, as a functional of $A_{\mu}^n$ and $B_{\mu \nu}$, has
to be (sufficiently) non-convex, which renders the effective action
"flat" in some region around the origin in field space. This "flatness"
corresponds to the above symmetry with all its consequences. This
phenomenon is evidently independent from the large N limit employed
here. The auxiliary scalar fields $\phi_i$ parametrize various bilinear
condensates which do not, however, break any internal symmetry.\par

We have argued -- but not shown in detail (to this end see \cite{14r})
-- that the present class of models can be obtained from $SU(N_c)$
Yang-Mills theories in the maximal abelian gauge after integrating out
the off-diagonal gluons. This application clarifies why it is sensible
to consider these models as effective low energy theories equipped with
an UV cutoff, but with a bare action including non-renormalizable
interactions. Our ansatz (\ref{1.3e}) for the bare action is already
quite general and exhibits the most interesting phenomena, but it could
easily be generalized by including further bilinear operators with
"internal" derivatives and/or external Lorentz indices. It can be
argued \cite{14r} that further  bilinear operators with "internal"
derivatives do not affect the low energy limit of the model (and just
modify somewhat the relation between the bare and effective actions),
whereas operators with external Lorentz indices will have no vevs (but
can affect the response of the model with respect to external sources).
More detailed investigations in this direction would be quite
straightforward.

A grain of salt constitutes the fact that the large N limit in the
present class of models does {\it not} coincide with the large $N_c$
limit of $SU(N_c)$ Yang-Mills theories \cite{14r}. First, the powers
of $N_c$ in $S_{bare}$ would not correspond to the powers of N
required in (\ref{1.5e}), and second $S_{bare}$ would not necessarily
depend only on $O(N)$ singlets since this $O(N)$ is not a sub-group of
$SU(N_c)$ (with $N_c = N + 1$). A discrete reflection symmetry, under
which $A_{\mu}^n$ and $B_{\mu \nu}^n$ (or the abelian components of 
$F_{\mu \nu}$) change sign is, however, a symmetry of $SU(N_c)$ 
Yang-Mills theories, which justifies at least the introduction of
bilinear composite operators. We recall again, on the other hand, that
the essential features of the confining phase do not rely on the large
N limit.\par 

We have seen that the "physics" of the confining phase is by no means
unique. On the one hand confinement can always be interpreted as
monopole condensation (and hence the vacuum as a dual superconductor),
but many features like the most important contributions to the string 
tension, (non-local) vacuum correlators and the shape of the vacuum
energy distribution perpendicular to a flux tube depend on the
non-universal properties of the model incoded in $S_{bare}$. Hence, if
we wish to learn more about the way confinement is realized in
Yang-Mills theories we have to find ways to learn more about the bare
action, and eventually to handle the present class of models beyond the
large N limit. Nevertheless the solvable version of the present models
can certainly play the role of a useful laboratory for the study of the
properties of a confining phase in the future. \par

\subsection*{Acknowledgement}

It is a pleasure to thank Y. Dokshitzer and A. Mueller for fruitful
discussions.

\vskip 5cm

\end{document}